\documentclass[aps,prl,twocolumn,linenumbers,superscriptaddress,showpacs, showkeys]{revtex4}
\usepackage{graphicx}
\usepackage{color}
\usepackage[english]{babel}
\usepackage{hyperref}

\begin{document}

\title{Thermolubricity of Xe monolayers on graphene}
\author{M. Pierno}
\affiliation{Dipartimento di Fisica e Astronomia ``G. Galilei'' Universit\'{a} di Padova
and CNISM\\
Via Marzolo 8, 35131, Padova, Italy}
\author{L. Bignardi}
\affiliation{Zernike Institute for Advanced Materials, University of Groningen\\
Nijenborgh 4, 9747AG Groningen, The Netherlands}
\author{M.C. Righi}
\email{mcrighi@unimore.it}
\affiliation{Istituto Nanoscienze, CNR and Dept of Physics, Informatics and Mathematics,
University of Modena e Reggio, 41125 Modena, Italy}
\author{L. Bruschi}
\affiliation{Dipartimento di Fisica e Astronomia ``G. Galilei'' Universit\'{a} di Padova
and CNISM\\
Via Marzolo 8, 35131, Padova, Italy}
\author{S. Gottardi}
\affiliation{Zernike Institute for Advanced Materials, University of Groningen\\
Nijenborgh 4, 9747AG Groningen, The Netherlands}
\author{M. St\"{o}hr}
\affiliation{Zernike Institute for Advanced Materials, University of Groningen\\
Nijenborgh 4, 9747AG Groningen, The Netherlands}
\author{P.L. Silvestrelli}
\affiliation{Dipartimento di Fisica e Astronomia ``G. Galilei'' Universit\'{a} di Padova
and CNISM\\
Via Marzolo 8, 35131, Padova, Italy}
\affiliation{DEMOCRITOS National Simulation Center, of the Italian Istituto Officina dei
Materiali (IOM) \\
of the Italian National Research Council (CNR), Trieste, Italy}
\author{P. Rudolf}
\email{p.rudolf@rug.nl}
\affiliation{Zernike Institute for Advanced Materials, University of Groningen\\
Nijenborgh 4, 9747AG Groningen, The Netherlands}
\author{G. Mistura}
\email{giampaolo.mistura@unipd.it}
\affiliation{Dipartimento di Fisica e Astronomia ``G. Galilei'' Universit\'{a} di Padova
and CNISM\\
Via Marzolo 8, 35131, Padova, Italy}
\date{\today}

\begin{abstract}
The nanofriction of Xe monolayers deposited on graphene was explored with a
quartz crystal microbalance (QCM) at temperatures between 25 and $50\;\text{K%
}$. Graphene was grown by chemical vapor deposition and transferred to the
QCM electrodes with a polymer stamp. At low temperatures, the Xe monolayers
are fully pinned to the graphene surface. Above $30\;\text{K}$, the Xe film
slides and the depinning onset coverage beyond which the film starts sliding
decreases with temperature. Similar measurements repeated on bare gold show
an enhanced slippage of the Xe films and a decrease of the depinning
temperature below 25 K. Nanofriction measurements of krypton and nitrogen
confirm this scenario.This thermolubric behavior is explained in terms of a
recent theory of the size dependence of static friction between adsorbed
islands and crystalline substrates.
\end{abstract}

\maketitle

Since its discovery, graphene has been found to possess numerous outstanding
properties such as extreme mechanical strength, extraordinarily high
electronic and thermal conductivity, thus opening the way to a plethora of
possible applications ~\cite{novoselov_roadmap_2012}. In particular, the
tribological features of graphene have received increasing attention in view
of the development of graphene-based coatings~\cite{lee_frictional_2010}.
Graphite is a well-known solid lubricant, used in many practical
applications. Its nanofriction behavior has been investigated mainly by
frictional force microscopy ~\cite%
{gnecco_fundamentals-book,Dienwiebel2004,koch_domes_2013}. Measurements on
few-layer graphene and single-layer graphene, prepared by micromechanical
cleaving on weakly adherent substrates, have revealed that friction
monotonically increases as the number of layers decreases~\cite%
{filleter_friction_2009,lee_frictional_2010,filleter2010}, while,
surprisingly, recent studies showed that this tendency is inverted when
graphene is suspended~\cite{Deng2012}.

Here we present the results of a quartz crystal microbalance (QCM) study
mainly focused on the sliding of \ Xe monolayers on graphene (Gr) between $20
$ and $50~\text{K, a temperature range }$which has been scarcely
investigated in the literature~\cite{danisman_simultaneous_2011}, despite
its relevance for the formation of condensed two-dimensional phases of many
simple gases\cite{bruch_progress_2007}. In our approach, the gold electrodes
of a QCM were covered with Gr because the ample availability of phase
diagrams of noble gases monolayers adsorbed on graphite~\cite%
{bruch_progress_2007} facilitates the interpretation of the QCM sliding
measurements~\cite{hosomi_sliding_2007,hosomi_dynamical_2009}.

In previous QCM experiments Gr was grown epitaxially on a Ni(111) QCM
electrode by heating the QCM to 400 $^{\circ }$C in the presence of carbon
monoxide~\cite{coffey_impact_2005,walker_frictional_2012}. However, no
direct morphological characterization of the resulting Gr coating was
reported. In our approach, Gr was transferred to the gold QCM electrode with
a polymer stamp and fully characterized with a variety of microscopies. Gr
was deposited by CVD on an ultra-pure copper foil (purity 99.999\%) in a
quartz-tube vacuum furnace (base pressure $10^{-5}$ mbar). The Cu foil was
reduced in H$_{2}$ (0.5 mbar) and Ar (0.1~mbar) for 60 min at 1180~K.
Subsequently Gr was grown by exposing the Cu foil to Ar (0.1~mbar), H$_{2}$
(0.5~mbar) and methane (0.5~mbar) for 2 min. at the same temperature~\cite%
{mattevi_review_2011}. Subsequently, the samples were cooled down to room
temperature in an Ar flux. Gr was then transferred onto a
polydimethylsiloxane (PDMS) stamp and the copper was etched away with an
aqueous solution of $\text{FeCl}_{3}$. After rinsing with milliQ water and
drying in a N$_{2}$ flow, the Gr layer was transferred from the PDMS stamp
onto the Au electrode of the quartz crystal by applying pressure and peeling
the stamp off.

\begin{figure}[tbp]
\includegraphics[width=\columnwidth]{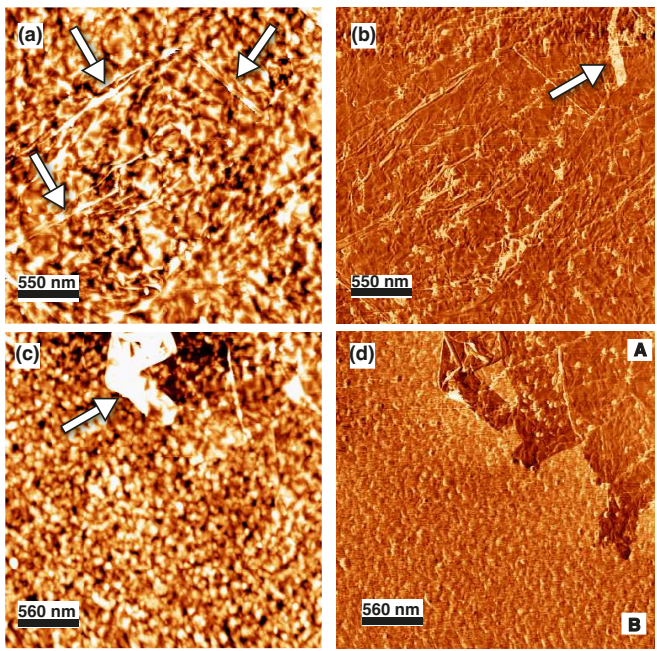}
\caption{(Color online) Topography (a) and lateral force (b) micrographs of
the graphene-coated quartz electrode collected with an atomic force
microscope. Brighter regions indicate higher height/friction. In the
topography image wrinkles on the surface (blue arrows) are evidence of the
presence of graphene. The lateral force micrograph shows homogeneous
friction all over the surface, except in correspondence of wrinkles and
holes in the graphene (blue arrow). Topography (c) and lateral friction (d)
micrographs of an area only partially covered by graphene close to the edge
of the gold electrode. The observed features and roughness are comparable to
those in panel (a). An area where the graphene is folded is indicated by the
arrow. In the lateral friction micrograph (d) region $A$ is covered by
graphene while in region $B$ the Au surface is bare.}
\label{fig:afm}
\end{figure}

The Gr layer on the Au electrode was characterized by contact-mode atomic
force microscopy (AFM). The transferred Gr covered approximately $90\pm 5\%$
of the electrode surface, as determined by a combination of AFM and optical
microscopy. Figure~\ref{fig:afm}-(a) shows a topography AFM micrograph ($%
3\times 3\;\mu \text{m}^{2}$) of a Gr-coated area where wrinkles in the Gr
layer are clearly visible (blue arrows); the root mean square (RMS)
roughness measured on such an area was $3.4\,\text{nm}$. Raman spectroscopy
(not shown) indicates that $95\pm 5\%$ is single layer Gr~\cite{lucaB-nano13}%
. Moreover, surface diffraction experiments carried out on layers prepared
in the same fashion yield sizes of single-crystalline grains ranging from
100~nm up to 5~$\mu \text{m}$~\cite{lucaB-nano13}. Figure~\ref{fig:afm}-(b)
presents a lateral force microscopy (LFM) scan of the same area shown in
Fig.~\ref{fig:afm}-(a), which appears homogeneous, except along the
wrinkles, where the lateral friction is higher. A hole in the Gr membrane
(blue arrow) appears as an area with higher friction. Figure~\ref{fig:afm}%
-(c) shows a topography AFM micrograph of an area at the electrode edge only
partially covered by Gr. The roughness of this area is uniform, revealing no
differences in topography between covered and uncovered areas, thus
suggesting that Gr is adhering to all asperities of the Au electrode. The
RMS roughness measured on bare Au is 2.6~nm. Finally, Fig.~\ref{fig:afm}-(d)
shows the LFM scan of the same area where two regions with different
friction are identified: the low-friction region A corresponds to the Gr
coating, the high-friction region B to bare Au, as explained in earlier
friction experiments on Gr~\cite{filleter_friction_2009}.

Identically prepared Gr-coated quartz crystals were inserted into a QCM
mounted\ and annealed to about $200{{}^\circ} \text{C}$ overnight in an UHV
chamber housing the cold head of a $4\;\text{K}$ cryocooler~\cite%
{bruschi_ultrahigh_2005}. Stainless steel spacers thermally decoupled the
QCM holder from the cold head. The adsorbate layer was condensed directly
onto the QCM, kept at the chosen low temperature, by slowly leaking
high-purity gas through a nozzle facing the quartz electrode. Between
consecutive deposition scans, the QCM was warmed up to about $60~\text{K}$
to guarantee full evaporation of Xe and thermal annealing of the
microbalance~\cite{pierno_nanofriction_2011}. The slip time $\tau _{s}$,
describing the viscous coupling between substrate and film, can be
calculated from the shifts in the resonance frequency and amplitude of the
QCM~\cite{bruschi_measurement_2001}. $\tau _{s}$ represents the time
constant of the exponential film velocity decay when the oscillating
substrate is brought to a sudden stop. Very low $\tau _{s}$ indicates high
interfacial viscosity; if a film is rigidly locked to the substrate, $\tau
_{s}$ goes to zero.

\begin{figure}[tbp]
\includegraphics[width=\columnwidth]{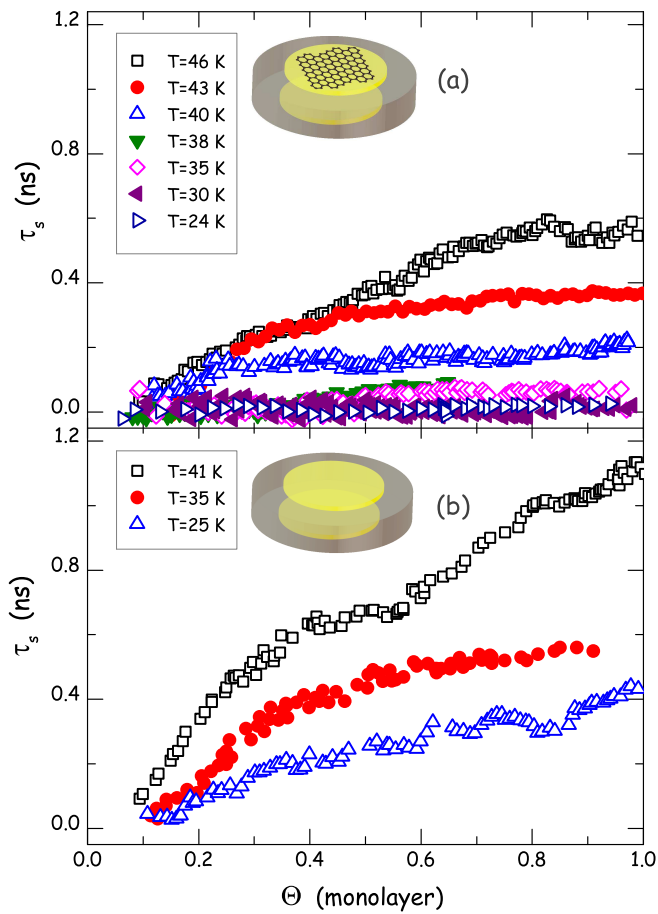}
\caption{(Color online) Slip time as a function of Xe coverage at different
temperatures for (a) Xe on graphene and (b) Xe on gold.}
\label{fig:xe}
\end{figure}

Figure~\ref{fig:xe}-(a) shows the slip time of Xe films deposited on Gr at
different temperatures, $\mathit{T}$, and for coverages $\Theta $ up to one
monolayer (ML). The coverage was deduced from the frequency shift assuming
for the ML an areal density of $5.94\;\text{atoms/nm}^{2}$, which
corresponds to the completion of a solid incommensurate phase on the
graphite lattice with nearest-neighbor distance $L_{nn}=0.441\;\text{nm}$~%
\cite{bruch_progress_2007,coffey_impact_2005}. This implies a frequency
shift for this ML of about $7.6\;\text{Hz}$. For each $T$, the average of a
few runs or the most representative scan is reported for the sake of
clarity. Data for $\Theta \leq 0.1$ ML are not plotted because of their
intrinsically large fluctuations.

At $T<30\;\text{K}$, $\tau _{s}$ is practically zero, indicating that the Xe
film is completely pinned to Gr. These findings are consistent with previous
studies at low $T$ reporting complete pinning of the highly polarizable $%
\text{N}_{2}$, Ar, Kr and Xe and sliding of the weakly polarizable Ne, $^{4}$%
He and $^{3}$He~on different surfaces.~\cite%
{renner_friction_2001,fois_low-temperature_2007,highland_superconductivity_2006,bruschi_structural_2006,pierno_nanofriction_2010,oda_prl13}%
.At $T=35\;\text{K}$, the film is initially pinned to the surface but starts
to slide for $\Theta >0.45$ ML. As the temperature is further increased, $%
\tau _{s}$ increases monotonically while the depinning onset coverage $%
\Theta _{\text{dep}},$ beyond which the film starts to slide, decreases
progressively. At the maximum temperature that could be achieved, $T=46\;%
\text{K}$, the slip time at monolayer completion is $\approx 0.5\;\text{ns}$
, much smaller than the value of $\approx 1.7\;\text{ns}$ measured at $77\;%
\text{K}$ \cite{coffey_impact_2005}. This behavior clearly suggests that the
sliding of the Xe film is favored by temperature. Similar thermolubric
effects have been reported for the friction of a tip moving along a graphite
surface~\cite{jinesh_thermolubricity_2008} and calculated in various models~%
\cite{vanossi_2013}. %

The slippage of Xe/Gr is significantly different from what observed on bare
Au electrodes. As shown in Fig.~\ref{fig:xe}-(b), the Xe slip times measured
on Au are much higher than those on Gr and $\Theta _{\text{dep}}$ is low
even at $T=25\;\text{K}$. This difference is more evident in Fig.~\ref%
{fig:thdep}, which reports the temperature dependence of $\Theta _{\text{dep}%
}$ for all systems. The error bars refer to the standard deviation of data
taken in different measurement runs. The vertical dashed line indicates the
temperature below which the Xe monolayer is found to be pinned to Gr. These
data are indicative of a thermal depinning transition with a characteristic
temperature $T_{\text{dep}}$ comprised between 30 and $35\;\text{K}$ for
Xe/Gr and below $25\;\text{K}$ for Xe/Au.

\begin{figure}[tbp]
\includegraphics[width=\columnwidth]{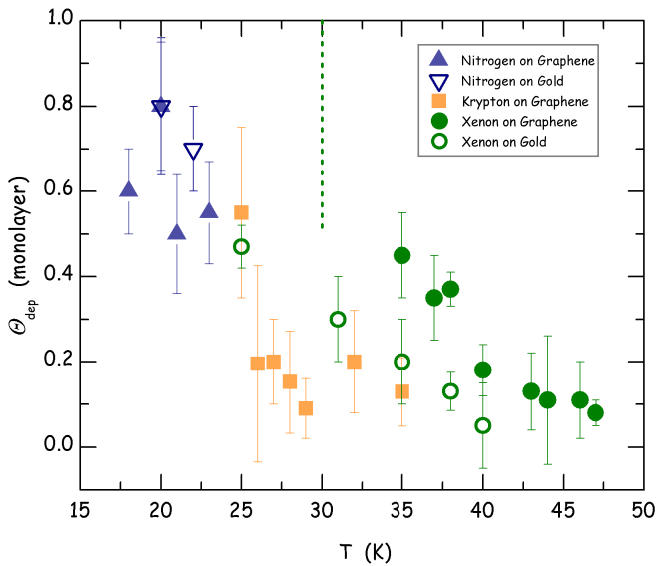}
\caption{(Color online) Depinning onset coverage as a function of
temperature for the systems investigated in this work. Dashed line indicates
the temperature below which Xe monolayers are always pinned to graphene.
Error bars account for data distributions over different runs performed at
different thermal cool downs from room temperature and different surfaces.}
\label{fig:thdep}
\end{figure}

As reported in Fig.~\ref{fig:thdep}, a similar trend was also observed for
Kr and $\text{N}_{2}$ films, the only difference being that the
characteristic temperatures are shifted to lower values reflecting the
smaller polarizability of these adsorbates as compared to Xe. For Kr/Gr, $%
\Theta _{\text{dep}}\approx 22\;\text{K}$ while, in the case of $\text{N}%
_{2} $, $\Theta _{\text{dep}}$ cannot be identified due to the narrow
temperature range accessible in our experiments. In fact, the data
acquisition is limited below, to about 20 K, by the thermal coupling of the
QCM to the head of the cryocooler and above, to $24\;\text{K}$, by the
evaporation of $\text{N}_{2}$ at $\Theta _{\text{dep}}\approx 1$. This
implies that $T_{\text{dep}} $ must be lower than $20\;\text{K}$ for both
surfaces.

One possible explanation of the observed difference between Gr and Au
surfaces is the higher corrugation of the surface potential on Gr with
respect to gold. Recent QCM experiments of Xe on a variety of metallic
electrodes have verified that the measured slip time is inversely
proportional to $U_{0}^{2}$, where $U_{0}$ is the amplitude of the periodic
function describing the changes in adsorbate-substrate potential with
respect to adsorbate position~\cite{coffey_impact_2005}. For Xe/Gr, $U_{0}$
amounts to $5.3\;\text{meV}$~\cite{coffey_impact_2005} while there is no $%
U_{0}$ value reported in the literature for Xe/Au. We have estimated the
corrugation amplitude $U_{0}$ for Xe and $\text{N}_{2}$ on Au and Gr using
an \textit{ab-initio} scheme based on the recently-developed nonlocal rVV10
density functionals~\cite{sabatini_nonlocal_2013} (including an accurate
description of van der Waals effects implemented in the QE package~ \cite%
{espresso}). The interaction of a Xe atom or a $\text{N}_{2}$ molecule with
both the ideal, planar single layer of Gr and the Au(111) surface were
considered. The computed $U_{0}$ value for Xe/Gr, $2.2\;\text{meV}$, is
significantly larger than that for Xe/Au Xe/Au ($1.6\;\text{meV}$), thus
supporting the explanation of the $\tau _{s}$ data based on the higher
corrugation of the surface potential on Gr than on Au. It is worthwhile to
point out that, although the absolute values of the \textit{ab-initio}
estimates of the corrugation could significantly depend on technical
details, such as the chosen density functional, their ratios are expected to
be much less sensitive to such particulars and therefore much more reliable.
Moreover the experimental estimate of $U_{0}$ obtained for Xe/Gr refers to
Gr grown on Ni(111)~\cite{coffey_impact_2005}, which could explain the
discrepancy with our value computed for an ideal, isolated layer of Gr. For $%
\text{N}_{2}$ the scenario is similar, since we find $U_{0}$ values of 8.0
and 6.6 meV for $\text{N}_{2}$/Gr and $\text{N}_{2}$/Au, respectively.

As for the observed temperature dependence of $\tau _{s}$, this is partly
consistent with dynamic simulations of the sliding of model Xe layers on
weakly corrugated surfaces~\cite{persson_theory_1993, vanossi_2013}. The
major difference between our data and the aforementioned molecular
simulations~\cite{persson_theory_1993} is the occurrence of a depinning
onset coverage which depends on temperature as clearly displayed in Fig.~\ref%
{fig:thdep} and which contrasts with the sliding at low $\Theta $ observed
in the simulations. %
%
A possible explanation for the observed decrease of $\Theta _{\text{dep}}$
with temperature relies on a recent theory~\cite{rapid12} about the size
dependence of static friction between adsorbed islands and crystalline
substrates according to which the atomic structure of islands deposited on a
substrate of nonmatching lattice parameters consists of commensurate domains
separated by incommensurate domain walls. Domain structures are governed by
the competition in minimizing both interfacial energy and elastic strain
energy. When the size of the contact is reduced below a critical radius, $%
R_{c}$, domains coalesce. This structural transition is accompanied by a
sharp increase of the interfacial commensurability and static friction~\cite%
{rapid12}. The depinning of a commensurate interface has been shown to be a
thermally activated process with an associated barrier $E_{\text{dep}%
}\propto \varepsilon U_{0}(F_{\text{s}}-F)/F_{\text{s}}F$, where $F$ is the
applied lateral force and $F_{\text{s}}$ the static friction force~\cite%
{pnas10}. In the case of perfect commensurability, the huge difference
between the static friction force and the weak inertia force provided by the
oscillating QCM results in a very high activation barrier (for example, $E_{%
\text{dep}}\simeq 10^{7}\;\text{eV}$ has been estimated for a Xe monolayer
on the Cu(111) surface in the absence of defects). Such barrier has been
found to drop dramatically by decreasing the interfacial commensurability~%
\cite{pnas10}. By combining these results with the experimental findings,
the following interpretation of the QCM data is proposed:

$i)$ The observation of a \emph{thermal activation of the frictional slip}
in nominally incommensurate Xe/Gr and Xe/Au systems can be accounted for by
the predicted increase of commensurability at small sizes. Islands of radius
lower than $R_{c}$ are expected to be pinned to the QCM, their thermal
depinning becoming probable at larger size due to the decrease of static
friction, $F_\textrm{s}$.

$ii)$ The \emph{critical coverage} necessary for the depinning of the film, $%
\Theta _{\text{dep}}$, is related to the coverage necessary for the growing
Xe islands to reach a large enough size, $R_{\text{dep}}\geq R_{c}$, for the
depinning process to be activated at the considered temperature. This seems
confirmed by the fact that the critical size estimated for Xe on Gr is
larger than that on Au %
in agreement with the experimental observation that $\Theta _{\text{dep}}$
is larger for Xe/Gr than for Xe/Au (see Fig.~\ref{fig:thdep}). The major
contribution to the calculated difference in the critical size comes from
the different misfit strain $e$. $R_{c}$ is, in fact proportional to $1/{%
e^{2}}$~\cite{rapid12}, where $e$ = -0.083 for Xe/Gr and $e$= 0.13 for Xe/Au~%
\cite{nota}.

$iii)$ The \emph{temperature dependence of the critical coverage} can be
accounted for by the two following arguments. By increasing the temperature, 
$\Theta _{\text{dep}}$ decreases because the probability to overcome the
barrier $E_{\text{dep}}$ increases, therefore the slip onset is activated
for higher values of $F_{\text{s}}$, that is for smaller island sizes.
Furthermore, $R_{\text{dep}}$ is reached at lower coverage due to enhanced
diffusivity at increased temperature.

This scenario applies not only to Xe/Gr but to all the systems investigated
in this work and should be quite general. Realistic temperature dependent
simulations of adsorbed films are clearly needed to better clarify the
phenomenon of thermal depinning observed in the experiments.

In summary, we have successfully managed to transfer graphene to the gold
electrode of a QCM and the adhesion is found to be good even at cryogenic
temperatures (e.g. down to 10 K). The presence of graphene on the gold
electrode significantly affects the sliding of Xe films. The measured slip
time is about half that on bare gold, probably because of the higher
corrugation of the surface potential on graphene with respect to gold.
Overall, the solid films are found to be rigidly pinned to the surface at
sufficiently low temperatures and start sliding at higher temperatures. The
onset coverage for sliding decreases with temperature and at a given
temperature is smaller on bare gold. Nanofriction measurements on krypton
and nitrogen confirm this scenario.This thermolubric behavior is explained
in terms of a recent theory of the size dependence of static friction
between adsorbed islands and crystalline substrates. This study provides the
first direct evidence of thermal lubricity of adsorbed films.

\begin{acknowledgments}
We thank Giorgio Delfitto and Luc Venema for invaluable technical
assistance, and Nanolab for the use of the clean-room. This work has been
partially supported by the Foundation for Fundamental Research on Matter
(FOM) in the framework of the \textquotedblleft Graphene-based
electronics\textquotedblright\ research program.
\end{acknowledgments}

\bibliographystyle{apsrev4-1}
\bibliography{graphene.bib}

\begin{thebibliography}{33}
\bibitem [{not()}]{nota}%
  \BibitemOpen
  \href@noop {} {}\bibinfo {note} {The misfit strain is calculated as $e = (a'
  -b)/b$, where $b$ is the lattice parameter of an unstrained Xe layer and $a'$
  is the parameter of the substrate sub-lattice occupied by the Xe atoms in a
  commensurate configuration. We considered a $\sqrt3\times\sqrt3$ sub-lattice,
  therefore $a' = \sqrt3 a$, with $a$ parameter of the hexagonal substrate
  surface.}\BibitemShut {Stop}%
\end{thebibliography}

\end{document}